\documentclass[12pt]{iopart}

\usepackage{iopams}  
\usepackage[pdftex]{graphicx}
\eqnobysec

\begin{document}

\title[Design and Construction of Desktop AC Susceptometer]{Design and Construction of a Desktop AC Susceptometer Using an Arduino and a Bluetooth for Serial Interface}
\author{Israel Perez$^1$, Jos\'e \'Angel Hern\'andez Cuevas$^2$ and Jos\'e Trinidad Elizalde Galindo$^2$}

\address{$^1$National Council of Science and Technology (CONACYT)-Institute of Engineering and Technology, Universidad Aut\'onoma de Ciudad Ju\'arez, Av. del Charro 450 Col. Romero Partido, C.P. 32310, Ju\'arez, Chihuahua, M\'exico}
\address{$^2$Institute of Engineering and Technology, Universidad Aut\'onoma de Ciudad Ju\'arez, Av. del Charro 450 Col. Romero Partido, C.P. 32310, Ju\'arez, Chihuahua, M\'exico}
\ead{cooguion@yahoo.com}
\begin{abstract}
We designed and developed a desktop ac susceptometer for the characterization of materials. The system consists of a lock-in amplifier, an ac function generator, a couple of coils, a sample holder, a computer system with a designed software in freeware C++ code, and an Arduino card coupled to a bluetooth module. The Arduino/bluetooth serial interface allows the user to have a connection to almost any computer and thus avoids the problem of connectivity between the computer and the peripherals such as the lock-in amplifier and the function generator. The bluetooth transmitter/receiver used is a commercial device which is robust and fast. These new features reduce the size and increment the versatility of the susceptometer for it can be used with a simple laptop. To test our instrument we performed measurements on magnetic materials and show that the system is reliable at both room temperature and cryogenic temperatures (77 K). The instrument is suitable for any physics or engineering laboratory either for research or academic purposes. 
\end{abstract}

\noindent{\it Keywords\/}: Susceptometer, Bluetooth, Magnetic Susceptibility:

\pacs{85.70.Ay, 85.25.Am}
\submitto{\EJP}
\maketitle
\section{Introduction}

Instruments for the determination of the ac magnetic susceptibility are essential in laboratories either in industry or research institutions. The measurement of the complex susceptibility of a sample is important because one can obtain information on the dynamic response of a material to an external magnetic field that otherwise cannot be assessed by conventional dc susceptometers. For several decades ac susceptometers have been routinely used in research because they analyzed the complex susceptibility revealing both the magnetic response of a material and magnetic losses caused by hysteretic behaviour. Their use has expanded to characterize a wide range of materials such as magnetic materials, superconductors and lately magnetic nanoparticles, bio- and nano-structured materials \cite{iperez10a,iperez10b,mfernandez15a,park09a,jklee03a}. In addition to this, they are used to characterize magnetic and phase transitions in ferromagnets, antiferromagnets, and superconductors \cite{gamboa15a,frolek08a}. On the other hand, headways made in the last two decades in informatics and electronics have not only favoured the accessibility of the public to more advanced and modern technologies but also have reduced the cost of these products. Nowadays more complex electronic devices and freeware are available in the internet at low cost and rapid accessibility. This has allowed researchers to develop more advanced projects, resulting in reduced size and higher efficiency. 

Taking advantage of these aspects herein we report on a simple design and construction of a desktop ac susceptometer. The instrument was designed to measure the dependence of the complex ac magnetic susceptibility $\chi=\chi'-i\chi''$ on the applied magnetic field $H$. This task is achieved using an ac function generator, a driving coil, a pair of pick-up coils coupled to a sample holder, and a lock-in amplifier. In turn these subsystems are controlled and monitored by a laptop. Finally, the whole system is automatized and the data are processed by a program run on C++ code which is freeware. One of the innovative aspects of this project resides in this freeware and the implementation of an interface via bluetooth. Most scientific instruments use either parallel general purpose interface bus (GPIB) or serial ports (e.g. RS-232, USB) to receive/transfer data from/to the computer. However, some types of connectivity are bulky, relatively expensive, and require special processing cards installed in computers (specially for GPIB connectivity) \cite{mfernandez15a,tafur12a}. To avoid these inconveniences we have implemented electronics based on an Arduino Uno coupled to a bluetooth module which allows reception/transmission to any modern computer via bluetooth. Since most laptops have bluetooth connectivity; this makes the system more versatile and compact, for there is no need for a desktop computer equipped with special cards. All of these improvements considerably reduce the size of the whole system, makes it more versatile and accessible to any physics or engineering laboratory.

\section{Theoretical aspects}
\label{theasp}
The theory of a susceptometer is based on the mutual induction between two coils, the driving coil and the pick-up coil \cite{nikolo95a}. Accordingly, the application of an ac current to the primary coil generates a sinusoidal magnetic field $H=H_{a0}\sin(\omega t)$, where $H_{a0}$ is the field amplitude, $\omega$ is the frequency $f$ multiplied by $2\pi$ and $t$ is the time. Subsequently, this field induces an electromotive force (emf) on the pick-up coil following Faraday's induction law
\begin{equation}
\label{femind1}
\epsilon(t)=-N\frac{d\Phi(t)}{dt}.
\end{equation}
Here $N$ is the number of turns of the pick-up coil and $\Phi(t)$ is the magnetic flux. If a material is introduced inside the coils the material will be magnetized and the flux can be expressed in terms of the sample magnetization $M(t)$ as $\Phi(t)=\mu_0AM(t)$; where $\mu_0$ is the vacuum permeability equal to $4\pi\times10^{-7}$ H/m and $A$ is the transversal area of the pick-up coils. In such case equation \eref{femind1} then becomes
\begin{equation}
\label{femind}
\epsilon(t)=-\mu_0AN\frac{dM(t)}{dt}.
\end{equation}
The magnetization is related to the field and the magnetic susceptibility by the relation $M=\chi H$, however, due to the magnetic response of the sample the magnetization is not sinusoidal and one can express the magnetization in a Fourier expansion as 
\begin{equation}
\label{eqmag}
M(t)=\sum_{n=1}^{\infty}H_{a0}\Big[\chi'_n\sin(n\omega t)-\chi''_n\cos(n\omega t)\Big];
\end{equation}
where $\chi'_n$ and $\chi''_n$ are the real and imaginary parts, respectively, of the harmonics of susceptibility $\chi_n=\chi'_n-i\chi''_n$ \cite{tishida90a}. So using  \eref{femind}, the emf becomes 
\begin{equation}
\label{eq1}
\epsilon(t)=\epsilon_0\sum_{n=1}^{\infty}n\Big[\chi'_n\cos(n\omega t)+\chi''_n\sin(n\omega t)\Big].
\end{equation}
For convenience we have made  $A=\pi a^2$ with $a$ the radius of the coil and $\epsilon_0=\mu_0 \pi a^2\omega NH_{a0}$. Using a lock-in amplifier the measured complex voltage ($\epsilon_n=\epsilon'_n-i\epsilon''_n$) can also be Fourier expanded and relate it to the susceptibility as
\begin{equation}
\label{voltsusc}
\epsilon'_n=\epsilon_0 n\chi'_n; \hspace{2cm} \epsilon''_n=-\epsilon_0n \chi''_n.
\end{equation}
Although this theory is general and thus higher harmonics of voltage may show up, we will restrict ourselves to work only with the first term of the series ($n=1$). More information on higher harmonics can be found elsewhere \cite{iperez10a}.

\section{Experimental setup}
\begin{figure}[t!]
\begin{center}
\includegraphics[width=6cm]{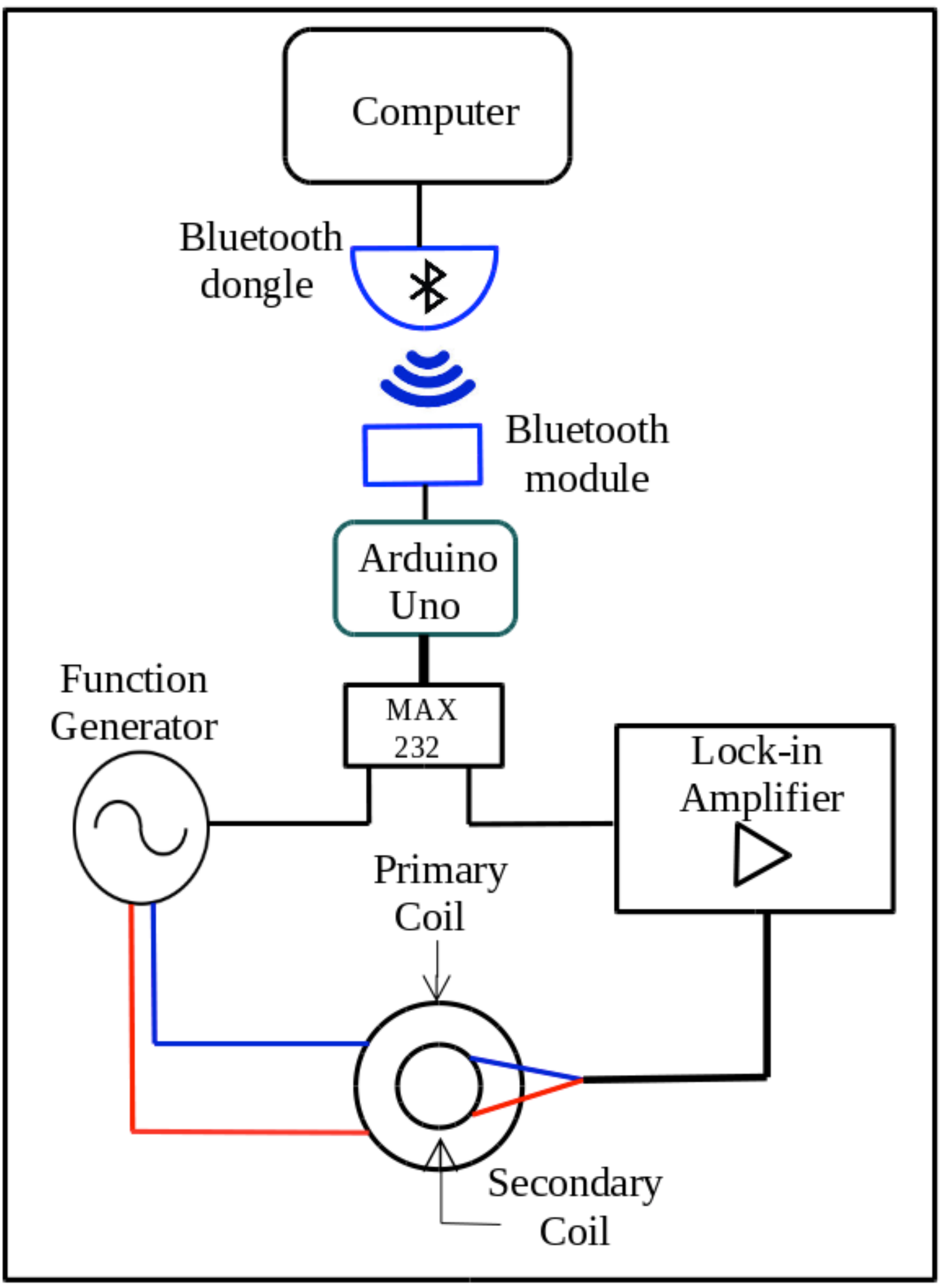}
\caption{Block diagram for the susceptometer.}
\label{diagramabloques}
\end{center}
\end{figure}

Our susceptometer is composed of several subsystems (figure \ref{diagramabloques}). Firstly, it has a primary coil (driving coil) which produces a maximum magnetic field of 3337 A/m at 1 kHz; a pair of secondary coils (pick-up coil) coaxially placed inside the driving coil. These coils are connected in series and wound in opposite directions in order to nullify the
induced voltage if no sample is inserted following Faraday's law equation \eref{femind}. Hence both coils have the same number of turns.
The sample is placed in a sample holder which in turn is inserted inside the lower pick-up coil. The primary coil is driven by an ac function generator (Frederiksen 2501.50) capable of supplying more than 0.5 A at frequencies below 5 kHz. A lock-in amplifier (Standford Research Systems, SR810) is used to measure the harmonics of the voltage induced in the pick-up coil. Most function generators are controlled by an RS-232 port with a DB-25 connector for serial interface while the amplifier has both GPIB and RS-232 ports, the later with a DB-25 connector. Taking advantage of Arduino Uno capabilities and commercial bluetooth modules we used the RS-232 ports and designed an electronic circuit to receive/transfer data via bluetooth between a laptop and these instruments. For this purpose we used the DB-25 connectors and transform the RS-232 signal ($\pm$12 V) used by the instruments to the TTL-signal used by the Arduino which in turn delivers the signal to the bluetooth module. This is a great advantage because DB-25 connectors can be acquired easily in comparison to GPIB connectors. Moreover, our system avoids special cards in computers and laptops which are required to manage either GPIB or DB-25 connectors.

\begin{figure}[t!]
\begin{center}
\includegraphics[width=8.5cm]{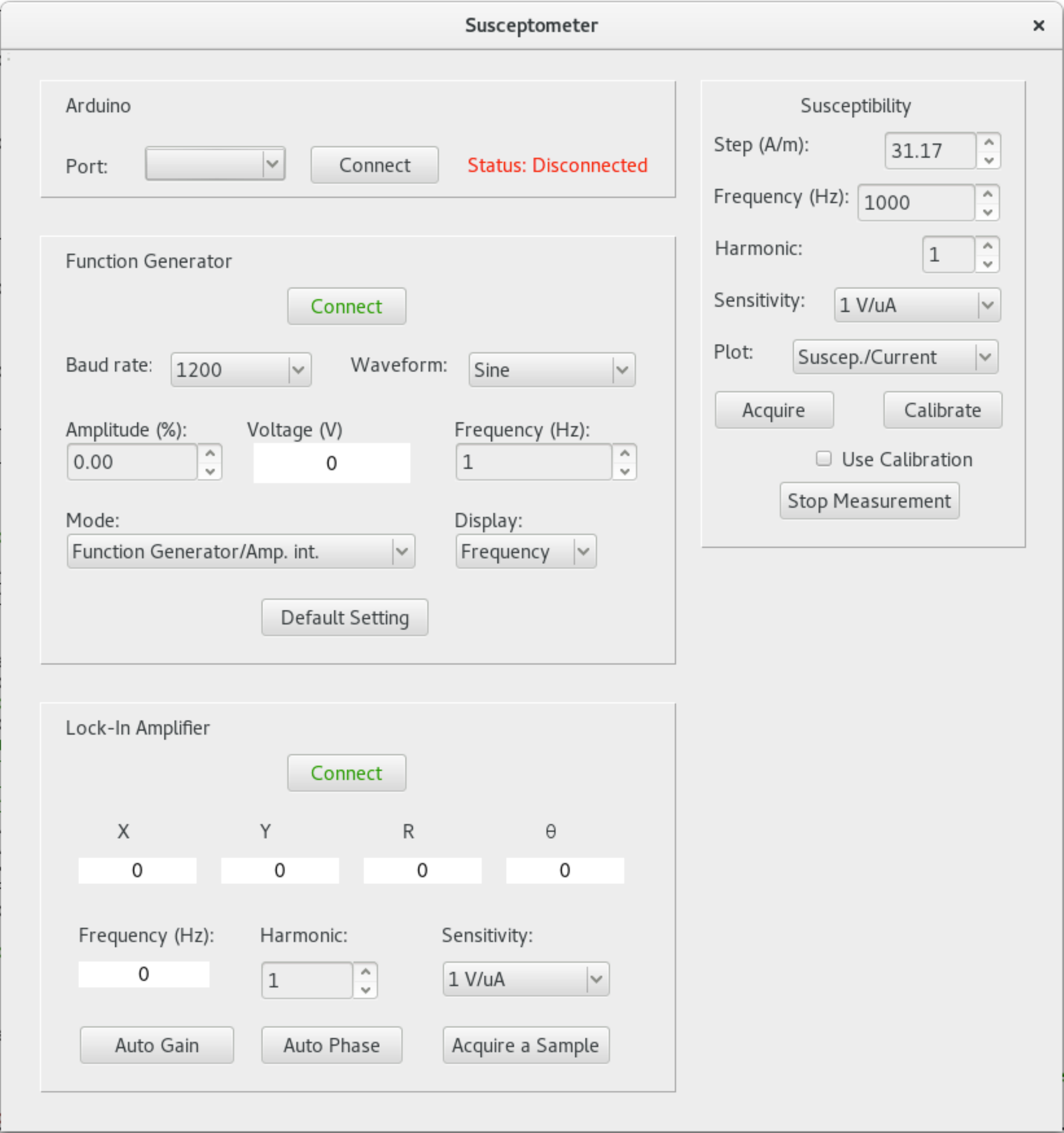}
\caption{Susceptometer graphical user interface.}
\label{interfaz2}
\end{center}
\end{figure}

To control and monitor the instruments we also designed a graphical user interface (GUI) in freeware C++ code. This is another important improvement since most GUIs are designed in LABVIEW code which is a commercial code \cite{mfernandez15a, park09a}. Therefore our project significantly reduces the cost of the system. The image of figure \ref{interfaz2} displays the GUI.
It is divided in mainly four parts, namely: 1) Arduino controller, 2) function generator, 3) Lock-in amplifier, and 4) Susceptibility. As the legends imply, the first part is used to establish communication with the Arduino. In the second part, one can control the function generator; especially the waveform, wave amplitude, and frequency. The third section was designed to manipulate the lock-in amplifier. Lastly the fourth section measures the dependence of the harmonics of susceptibility on the applied magnetic field. A calibration button is required to ensure the induced voltage in the pick-up coils be nullified; that is, before running a measuring session with a sample, the system is calibrated without a sample. One depresses this button and a measurement is run to determine the offset voltage. This signal is stored for future reference and in subsequent measurements with samples, the software retrieves it and subtracts it from the last measurement so the reported signal strictly depends on the susceptibility of the sample.

\section{Coil design}
\begin{figure}[t!]
\begin{center}
\includegraphics[width=6.5cm]{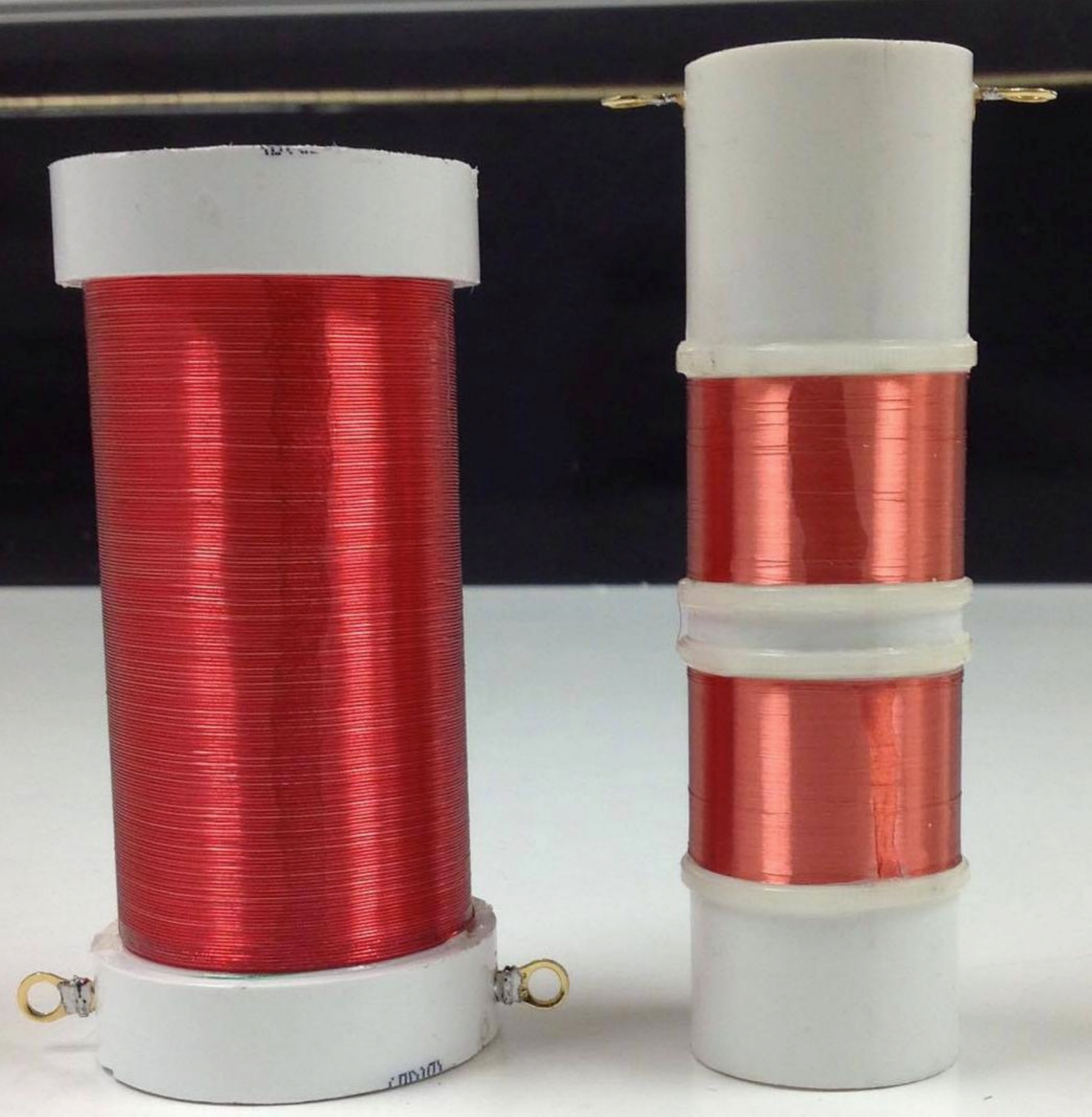}
\caption{Driving coil (left) and pick-up coil (right).}
\label{fcoils}
\end{center}
\end{figure}

We have designed two coils, the primary and the secondary coils. These coils are shown in figure \ref{fcoils} and their corresponding data is given in table \ref{tcoils}.
The dimensions of the coils were mainly based on the sample size to be used. Since it is desirable to have a homogenous field all over the sample, it is required that the dimensions of the sample are small enough to be considered as a small dipole, so a sufficiently wide and long driving coil is needed. In order to meet this condition we restricted the sample thickness to a maximum of 5 mm which represents only 6\% of the coils' length. The sample diameter is only limited by the sample holder which in this case is about 25 mm. This has allowed us to characterize samples in different presentations such as chunks, pellets, and even powder that can be previously encapsulated in small non-magnetic containers. 
\begin{table}[b!]
  \centering 
  \caption{Data for the coils.}
  \label{tcoils}
  \begin{tabular}{c|c|c}
\hline
\hline Feature & Driving coil  & Pick-up coil  \\ \hline
  Number of turns & 736  &  278 each \\
  Wire thickness  & 26 AWG  & 40 AWG  \\ 
 Coil length/mm & 81.0 & 25 each \\
 External radius/mm & 21.2 & 16.5  \\
Internal radius/mm  & 19.0 & 15.0 \\
\hline
\end{tabular}
\end{table}

To assess the magnetic field generated by our coils, we performed a simulation and verified the magnitude of the magnetic field carrying out some measurements. The theory used for the simulation is based on the Biot-Savart law that governs the magnetic field $\vec{B}$ generated by an element of length $d\vec{l}$ pointing in the direction of current flow \cite{young16a}:
\begin{equation}
d \vec{B}= \frac{\mu_{0}}{4 \pi} \frac{I d\vec{l} \times \hat{r}}{r^{2}},
\label{equ:biot-savart}
\end{equation}
where $r$ is the distance from the element to the point of interest (see below). For our calculations we have defined the origin of coordinates at the centre of the coil and the $z$-axis pointing parallel to the coil axis. For a coil of $N$ turns, length $\ell$ and radius $a$ the element of length is:
\begin{equation}
d\vec{l}(\theta)=-a \sin(\theta) \hat{\i}+a \cos(\theta) \hat{\j}+ \frac{\ell }{2\pi N}\hat{k} \qquad  \textrm{for} \quad 0 \le \theta \le 2\pi N.
\end{equation}
Now we can define the vector $\vec{r}$ as the vector that goes from a point in the coil to an arbitrary point $(x,y,z)$ in space and its direction as
\begin{equation}
\hat{r}= \frac{b}{r} \hat{\i} + \frac{c}{r} \hat{\j} +\frac{d}{r} \hat{k};
\end{equation}
where $b=x-a \cos(\theta)$, $c=y-a \sin(\theta)$, $d=z-\frac{\ell}{2}\Big(\frac{\theta}{N\pi}+1\Big) $, and $r=\sqrt{b^{2}+c^{2}+d^{2}}.$
\begin{figure}[b!]
\begin{center}
\includegraphics[width=15cm]{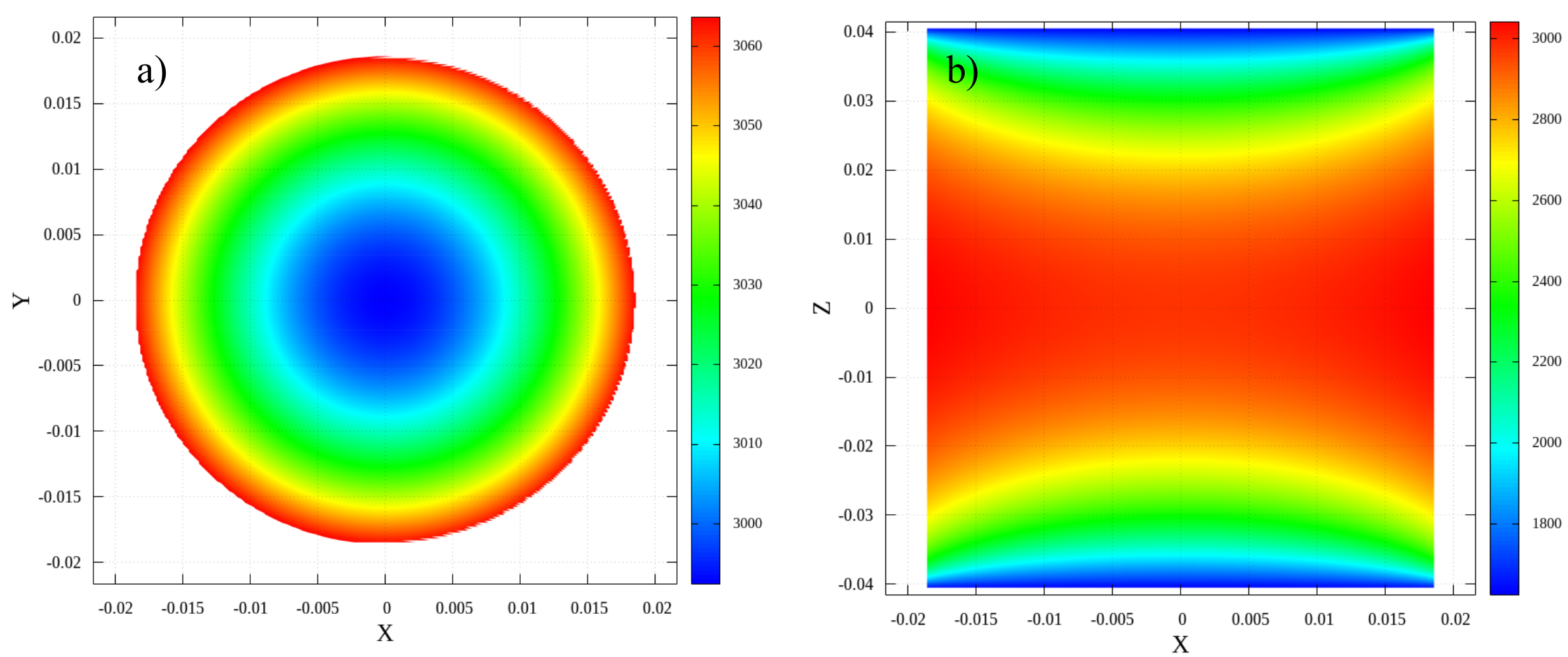}
\caption{$H_z$ component of the magnetic field for the driving coil. a) Plane $x-y$ with $z=0$ and b) Plane $x-z$ with $y=0$. The colour scale is given in A/m and the coordinate axes are in m.}
\label{campobz}
\end{center}
\end{figure}

For our convenience, we will focus on the component along the coil axis ($B_z$) because the others have much smaller magnitudes and thus play a minor role. So, after performing the cross product $d\vec{l}(\theta) \times \hat{r}$ and dealing with cumbersome algebra, we arrived at:
\begin{equation}
\label{bz}
H_z=\frac{B_{z}}{\mu_{0}}=-\frac {I}{4 \pi} \int_0^{2 \pi N} \frac{a}{r^3}\left[c\sin(\theta)  +b\cos(\theta)\right] d\theta.
\end{equation}
The solution to this equation was found by numerical integration; results are shown in figure \ref{campobz}(a) for the plane $x-y$ and figure \ref{campobz}(b) for the plane $x-z$. In the former case we can see that the field at the centre of the coil is homogeneous so long as the sample dimensions do not exceed half of the radius ($a=1.9$ cm). According to these results, it is therefore desirable to place a small sample along the coil axis. At the origin of coordinates ($x=y=z=0$ cm) the field for a current of 745 mA is about 3000 A/m and at $x=y=a$ the field is 3060 A/m. By comparison the measurements for the origin give 3000.1 A/m and at $x=y=1.9$ cm is 3055.8 A/m in excellent agreement with the theory. Regarding the homogeneity along the coil axis, the graph for the plane $x-z$ shows that the field does not vary much for an area around the origin of 2 cm$^2$. For instance, the measured field at $z=1$ cm and $x=y=0$ cm is 2936.4 A/m and at $z=1$ cm, $x=1.9$ cm, and $y=0$ cm is 2984.1 A/m; whereas the calculated field in these same positions is 2930 A/m and 3000 A/m, respectively. And since the sample thickness is less than 1 cm, we note that the variation in field intensity amounts to just about 1\%.

We also evaluated the magnetic field response of the driving coil to variations in frequency for several applied voltages, namely: (1, 3, 5, 7) V. We observed that the field intensity decreases as a function of frequency. 	
\begin{figure}[t!]
\begin{center}
\includegraphics[width=8.5cm]{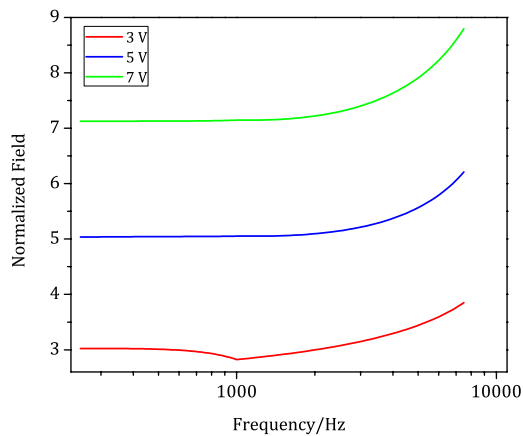}
\caption{Frequency dependence of the magnetic field generated by the driving coil at different applied voltages. The curves have been normalized relative to the data at 1 V.}
\label{HvsFreq}
\end{center}
\end{figure}
This is expected since the impedance $\zeta$ of the coils depends on both the self-inductance $L$ and the angular frequency through the equation $\zeta=\sqrt{R^2+(\omega L)^2}$. Thus, as $\omega$ increases the impedance increases and the current in the coils reduces; and since the field intensity is proportional to the current $I$, the field amplitude also decreases. This effect is not desirable when performing measurements as a function of frequency. This however does not affect our results since we carried out our experiments in a fixed frequency. In figure \ref{HvsFreq} we display the dependency of the normalized field on the frequency. The curves have been normalized with respect to the data for 1 V. There we can observe that the field remains almost constant for frequencies smaller than 1.5 kHz and increases for higher frequencies. This indicates a nonlinear dependence of the field with respect to the applied voltage for relatively high frequencies.

\section{Electronics}
The electronics is another important part of the susceptometer. As mentioned above the system is based on an Arduino Uno and a bluetooth module. The bluetooth module is used for transmitting/receiving the signal between the computer and the instruments (lock-in amplifier and function generator). The coupling between the instruments and the bluetooth module is realized by means of the Arduino. In this way one can use a bluetooth module (or dongle) in the computer and control the system remotely; thus avoiding wire connections.
\begin{figure}[t!]
\begin{center}
\includegraphics[width=8.5cm]{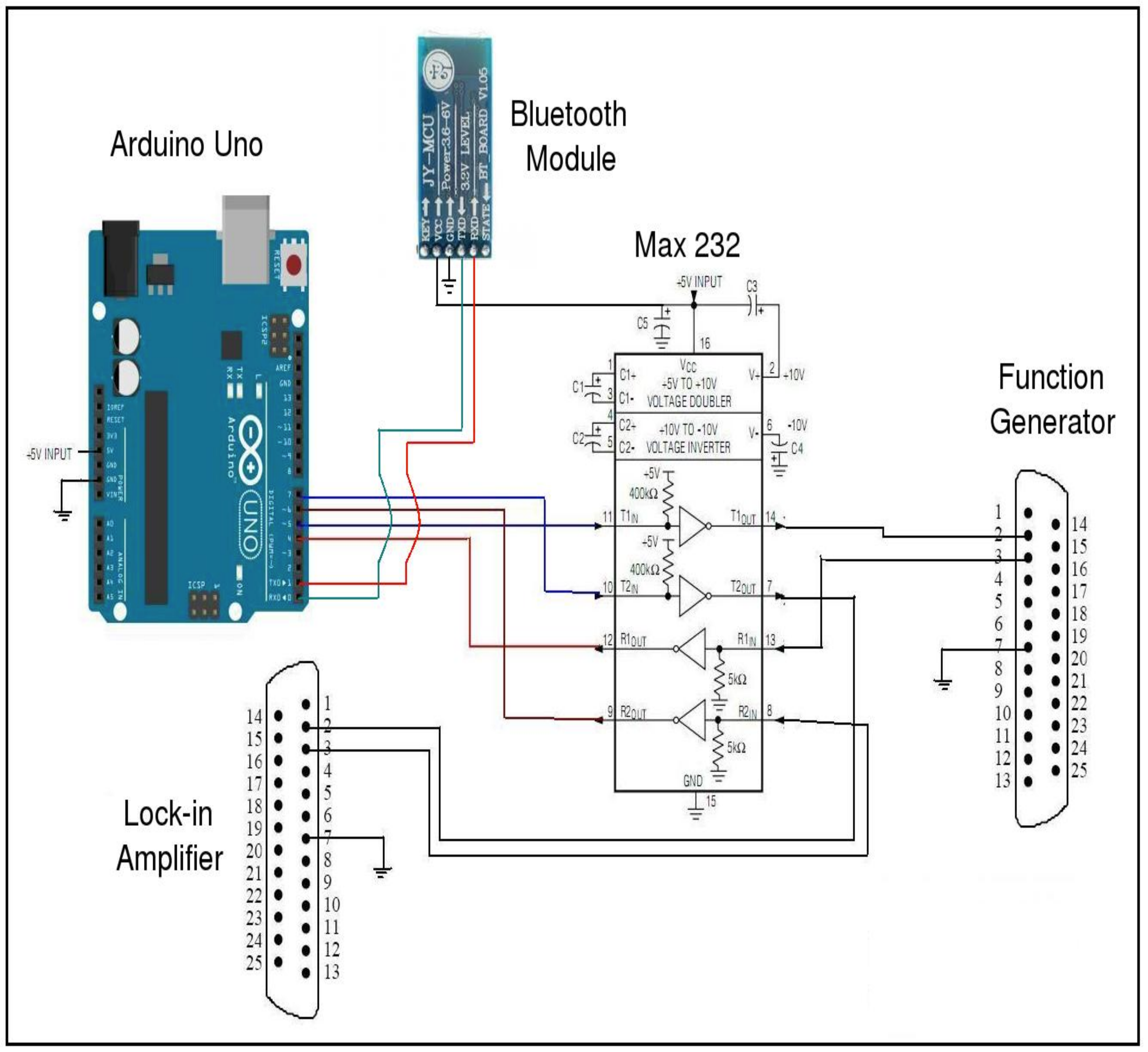}
\caption{Interface between the laptop and the instruments. An Arduino Uno is coupled to the instruments via DB-25 cables and a MAX232 IC.}
\label{max232arduino2}
\end{center}
\end{figure}
The instruments are connected to the Arduino by two serial ports RS-232 through DB-25 connectors that transmit signals in the voltage range $\pm$12 V, but the Arduino works with TTL logic. So we need a device capable of converting the $\pm$ 12 V signal to 0 V to 5 V signal. This conversion is achieved by an integrated circuit MAX232 (see figure \ref{max232arduino2}) that converts -12 V to 5 V and +12 V to 0 V (more details of the electronic circuit are given in the appendix).

\section{Frequency and instrument calibration}
\subsection{Frequency}
We just discussed that there is a strong dependency of the frequency on the magnetic field of the driving coil and that higher harmonics of susceptibility may show up (this mostly depends on the magnetic properties of each material); thus with the aim of avoiding the generation of higher harmonics and the degradation of the field amplitude at high frequencies, for our experiments we set the frequency to 1 kHz. This is one of the most typical values reported in the literature and it is therefore helpful for comparison purposes \cite{iperez10a,iperez10b}.

\subsection{Calibration factor and sensitivity}

The sensitivity of our susceptometer depends on several factors such as sample volume $V$, lock-in amplifier resolution, frequency, applied field, etc., and to estimate its value we must determine a constant known as the calibration or filling factor $\alpha$. In section \ref{theasp} we outlined the theory that relates the induced voltage $\epsilon$ to the susceptibility $\chi$. In particular, equation \eref{eq1} is strictly true for a uniformly magnetized sample; this in practice however does not occur. Thus, in general $\epsilon$ must depend on the geometry and spatial orientation of the sample and the pick-up coil and such information is contained in the filling factor. So, according to our experimental setup, the lock-in amplifier measures the root mean square (rms) value of the induced voltage $\epsilon$, which is proportional to the susceptibility following the relation \cite{nikolo95a}:
\begin{equation}
\label{factor}
|\chi|=\alpha \frac{\epsilon{\tiny{\textrm{rms}}}}{VfH_{\tiny{\textrm{rms}}}}.
\end{equation} 
The calibration factor can be estimated either from geometrical considerations or by using a standard material of known susceptibility such as a superconductor \cite{nikolo95a,couach92a,malderighi13a}. In our case, we have decided to use an YBa$_2$Cu$_3$O$_{7-\delta}$ (YBCO) superconductor purchased from Colorado Superconductors. As is well known, superconductors are the only materials that exhibit perfect diamagnetism ($\chi=-1$), a property exclusively manifested during the Meissner-Ochsenfeld state; and therefore, in this state, the calibration factor can be determined. 

Before calibration we verified this state by introducing the superconductor in a liquid nitrogen bath at 77 K. This temperature is enough to drive the material into the superconducting state since the transition temperature of bulk YBCO is 90 K. The manifestation of the Meissner-Ochsenfeld effect not only implies that the material is in the superconducting state but also that $\delta$ is between 0.1 and 0.3; indicating an optimal non-stoichiometry. According to some reports the lower critical field for YBCO, with $\delta$ between 0.1 and 0.3, is between 4000 A/m and 8000 A/m at 77 K, respectively \cite{liang94a,bohmer97a}. Since the driving coil generates less than 3500 A/m, the superconductor must be, at least as the field is initially increased, in the Meissner-Ochsenfeld state; meaning that for the initial applied field $\chi'\approx-1$; as required.
\begin{figure}[t!]
\begin{center}
\includegraphics[width=9.5cm]{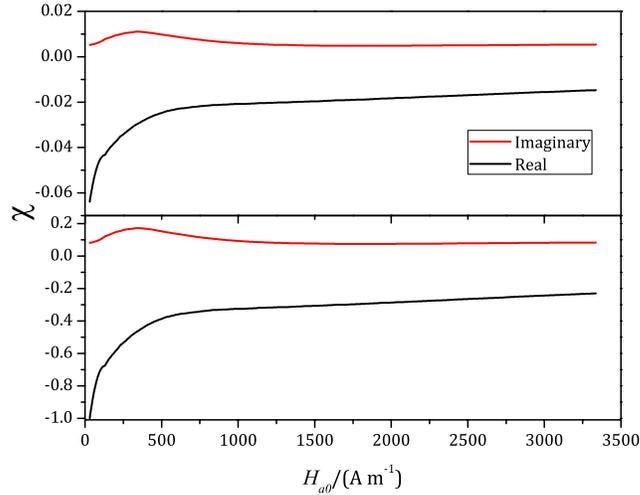} 
\caption{Susceptibility for a YBa$_2$Cu$_3$O$_{7-\delta}$ superconductor at 77 K. Plots for raw data (top) and after correction from calibration factor (bottom).}
\label{sc1077K}
\end{center}
\end{figure}
Afterwards we carried out a measurement of susceptibility for this sample and found a calibration factor of 15.64. Figure \ref{sc1077K} shows the results of the measurements before and after the calibration correction. As we can notice the shape of the real and imaginary components are in excellent agreement with both experiment and theoretical calculations for YBCO, thus supporting the reliability of our system \cite{iperez10a}.

With the calibration factor determined, we can compute the sensitivity of our instrument using equation \eref{factor} for a voltage resolution of 1 nV and a typical volume of 2 cm$^3$; in such case the sensitivity for a signal of 1 kHz is $|\chi|=7.8\times10^{-6}$. 

\section{Applications}
\begin{figure}[t!]
\begin{center}
\includegraphics[width=9.5cm]{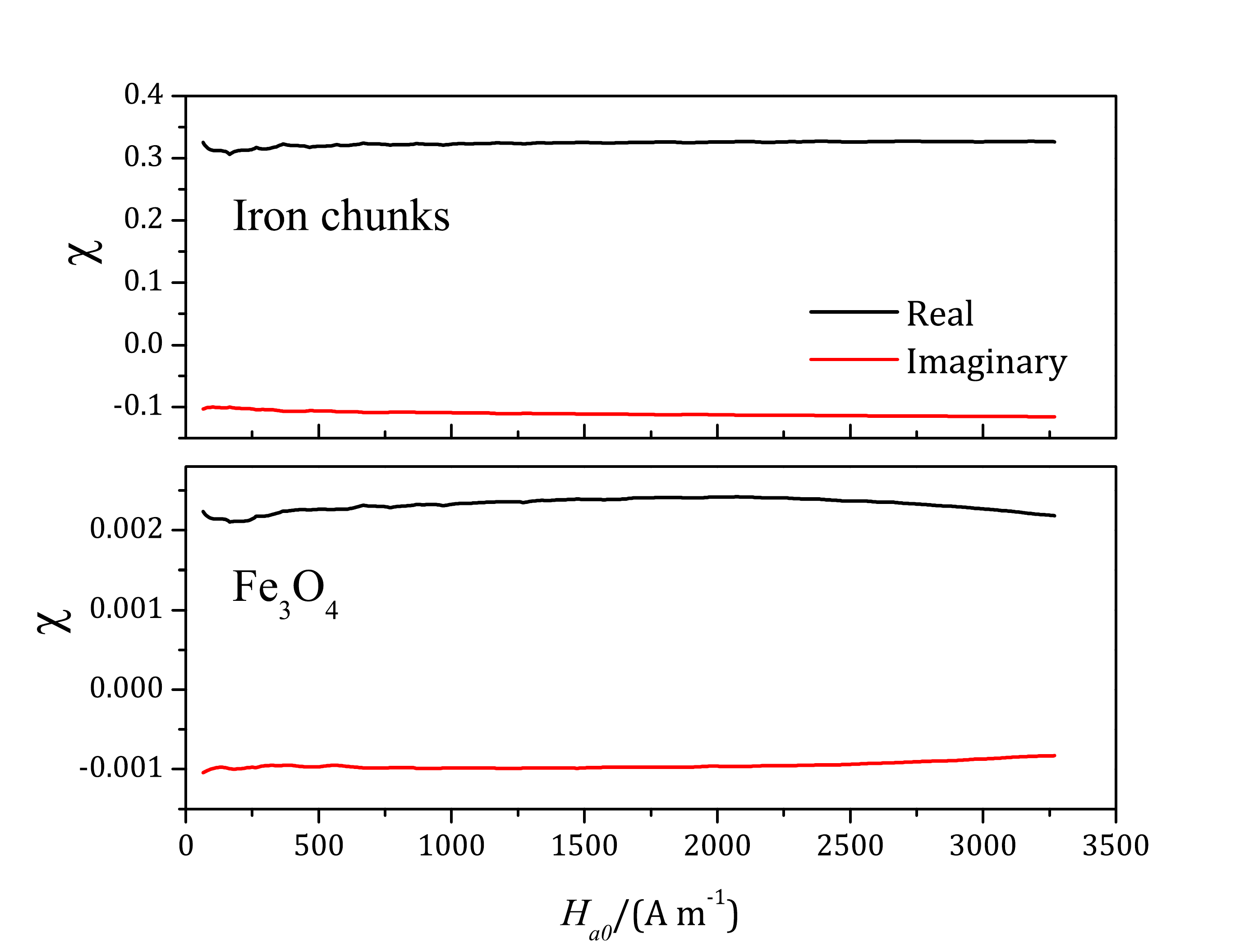} 
\caption{Susceptibility for iron chunks (top) and magnetite nanoparticles (bottom) both at 300 K.}
\label{fefe3o4}
\end{center}
\end{figure}

The susceptometer is versatile and can be used to characterize materials of any class so long as they meet the size requirements and these include materials in chunk, pellets, or micro- and nano-particles (previously encapsulated in non-magnetic containers). To test the reliability of our system we measured the susceptibility of iron chunks and magnetite nanoparticles (Fe$_3$O$_4$) at room temperature and the results are shown in figure \ref{fefe3o4}; we note that prior to and during the measurements the samples were not exposed to dc magnetic fields, so their dc magnetization is zero. 
There we can observe that the susceptibility of iron is constant as the magnetic field increases. In comparison, the susceptibility of magnetite slightly varies as the field rises and its magnitude is two orders of magnitude smaller than its counterpart. We also see that the real components of both samples are positive; indicating the magnetic character of the samples just as expected and in good agreement with previous reports \cite{mjackson98a,fhrouda02a}. Similarly, the imaginary components are negative; indicating the energy losses due to hysteretic behaviour typical in ferro and ferrimagnets. 

\section{Conclusions}
\label{con}
We successfully built an ac susceptometer to determine the dependence of the complex ac susceptibility of samples on the applied magnetic field. The instrument generates fields up to 3337 A/m and is capable of working in temperatures between 77 K and room temperature. The system has some improvements regarding the transmission/reception of data from the instruments to the computer based on an Arduino and a bluetooth module, which serve to control the system remotely. The susceptometer is reliable since the calibration of the system was realized by means of a superconductor taking the condition $\chi=-1$. We verified that the shape and magnitude of the curve of susceptibility as function of applied field for a our superconductor is in excellent agreement with the results found in the literature. We also applied our instrument to determine the complex susceptibility at room temperature of iron chunks and magnetite nanoparticles. The results are coherent with regards to the magnitude, shape, and sign of the real and imaginary components and in agreement with those reported in the literature. This consequently supports the reliability of our instrument.

\ack
We are grateful to Dr Karen Castrejon for her support in the physics lab. The authors gratefully acknowledge the support from the National Council of Science and Technology (CONACYT) Mexico and the program C\'atedras CONACYT through project 3035. We are also indebted to the anonymous reviewers for their critics and comments that greatly improved the quality of this work.

\appendix

\section{Electronic circuitry}
\begin{figure*}[t!]
\begin{center}
\includegraphics[width=17cm]{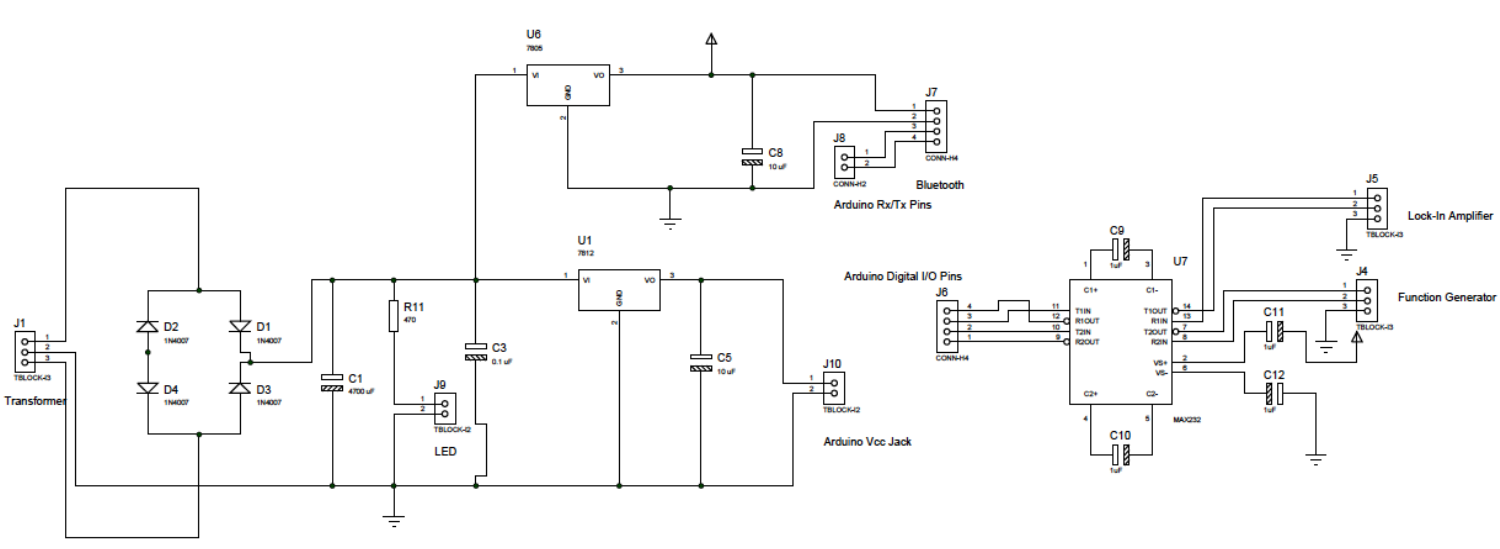}
\caption{Electronic schematic diagram.}
\label{schemdiagram}
\end{center}
\end{figure*}

\begin{figure*}[t!]
\begin{center}
\includegraphics[width=15.7cm]{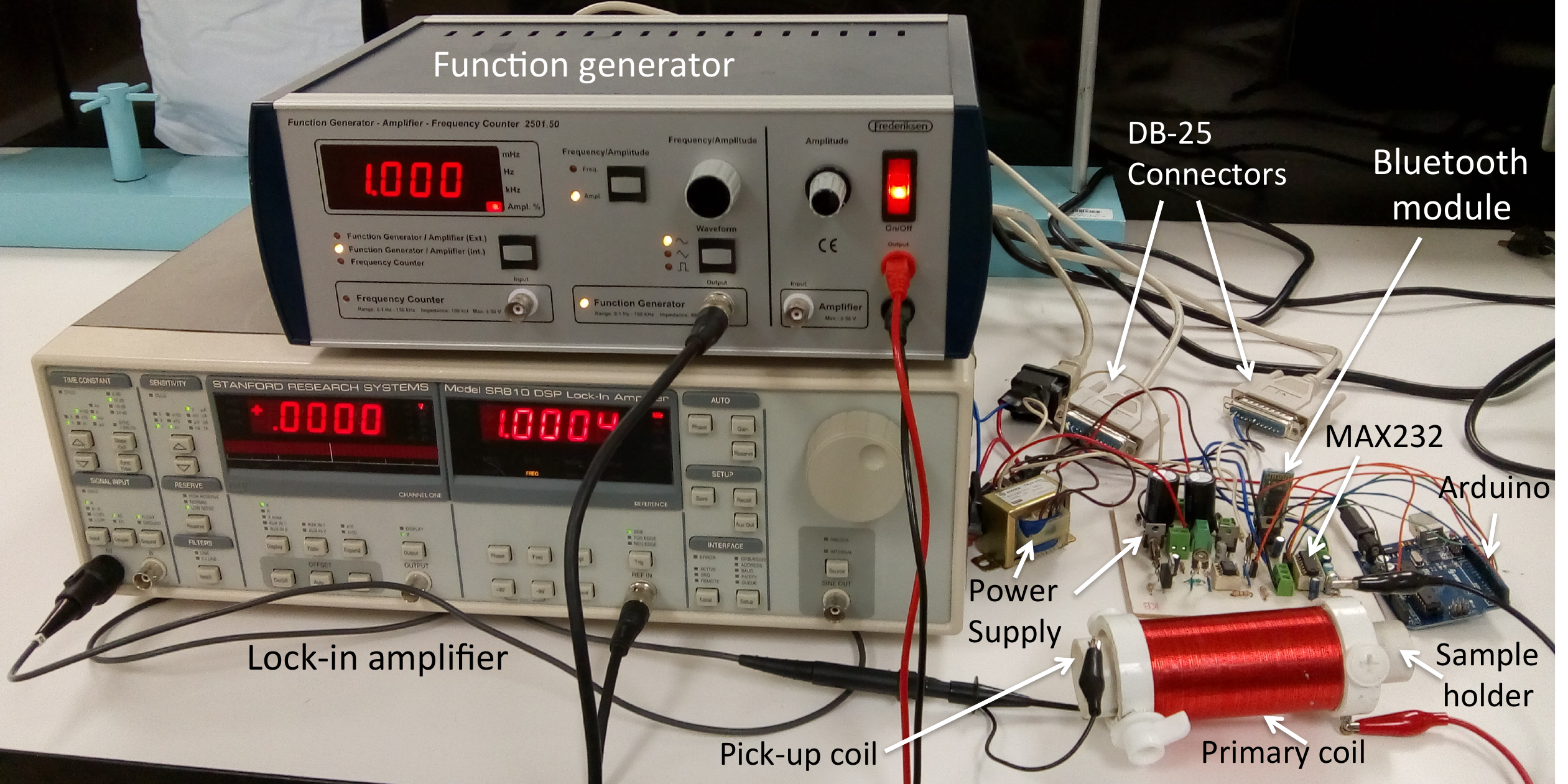}
\caption{Instrumentation assembly.}
\label{9378}
\end{center}
\end{figure*}

The electronic circuitry is integrated in a single home-made printed circuit board specially designed to fulfill our requirements. The whole schematic diagram is displayed in figure \ref{schemdiagram} and a picture of the instrument assembly is shown in figure \ref{9378}.
The circuits are fed by a power supply with a 1 A-rectifier bridge. The maximum current is 500 mA with dc output voltages of 5 V and 12 V. The first supply is used to feed both the MAX232 and the bluetooth module, while the second one goes to the Arduino card.


\section*{References}

\begin{thebibliography}{99}

\bibitem{iperez10a} Perez I, Gamboa F, and Sosa V 2010 Study of higher-order harmonics of complex ac susceptibility in YBa$_2$Cu$_3$O$_{7-x}$ thin films by the mutual inductive method  {\it Physica C} {\bf 470} 2061

\bibitem{iperez10b} Perez-Lopez I, Gamboa F, and Sosa V 2010 Critical current density and ac harmonic voltage generation in YBa$_2$Cu$_3$O$_{7-x}$ thin films by the screening technique {\it Physica C}  {\bf 470}, S972

\bibitem{mfernandez15a} Fernandez M P, Teixeira J M, Machado P, Oliveira M R F F, Maia J M, Pereira C, Pereira A M, Freire C, and Araujo J P, 2015 Automatized and desktop AC-susceptometer for the in situ and real time monitoring of magnetic nanoparticlesÕ synthesis by coprecipitation {\it Rev. Sci. Instrum.} {\bf 86} 043904

\bibitem{park09a} Park K, Sonkusale S, Guertin R P, Harrah T and Goldberg E B 2009 A miniaturized AC magnetic susceptometer for detecting biomolecules tagged to magnetic nanoparticles {\it Bioengineering Conference} doi:10.1109/NEBC.2009.4967761

\bibitem{jklee03a} Lee J K and Choi S M 2003 {\it Bull. Korean Chem. Soc.} {\bf 24} 32

\bibitem{gamboa15a} Gamboa F, Perez I, Matutes-Aquino J A, Moewes A and Sosa V 2015 Determination of the Critical Current Density in YBa$_2$Cu$_3$O$_{7-x}$ Thin Films Measured by the Screening Technique Under Two Criteria {\it IEEE Trans. on Appl. Supercond.} {\bf 25} 8000105

\bibitem{frolek08a} Frolek L, G\"omory F and Seiler E 2008 A special sample holder for AC susceptibility measurements of superconducting samples in high magnetic fields at various temperatures {\it Supercond. Sci. Technol.} {\bf 21} 105009

\bibitem{tafur12a} Tafur J, Herrera A P, Carlos R and Juan E 2012 Development and validation of a 10 kHz-1 MHz magnetic susceptometer with constant excitation field {\it J. Appl. Phys.} {\bf 111} 07E349

\bibitem{nikolo95a} Nikolo M, 1995 Superconductivity: A guide to alternating current  susceptibility measurements and alternating current susceptometer design {\it Am. J. Phys.} {\bf 63} 57

\bibitem{couach92a}  Couach M and Khoder A F 1992 {\it Magnetic Susceptibility of Superconductors and Other Spin Systems}, Eds. Francavilla T L, Hein R A and Liebenberg D H (NY: Plenum).

 \bibitem{young16a} Young H D and Freedman R A 2016 {\it University Physics with Modern Physics}, 14th edn (NJ: Pearson)
 
  \bibitem{malderighi13a} Alderighi M, Bevilacqua G, Biancalana V, Khanbekyan A, Dancheva Y and Moi L 2013 A room-temperature alternating current susceptometer -- Data analysis, calibration, and test {\it Rev. Sci. Instrum.} {\bf 84} 125105.
  
  \bibitem{tishida90a} Ishida T and Goldfarb R B 1990 Fundamental and harmonic susceptibilities of YBa$_2$ Cu$_3$O$_{7-\delta}$ {\it Phys. Rev.} B {\bf 41} 8937

\bibitem{liang94a} Liang R, Dosanjh R , Bonn D A, Hardy W N and Berlinsky A J 1994 Lower critical fields in an ellipsoid-shaped YBa$_2$Cu$_3$O$_{6.95}$ single crystal {\it Phys. Rev. B.} {\bf 50} 4212
 
 \bibitem{bohmer97a} B\"ohmer C, Brandst\"atter G and Weber H W 1997 The lower critical field of high-temperature superconductors {\it Supercond. Sci. Technol.} {\bf 10} A1

\bibitem{mjackson98a} Jackson M, Moskowiiz B, Rosenbaum J and Kissel C 1998 Field-dependence of AC susceptibility in titanomagnetites {\it Earth Planet. Sci. Lett.} {\bf 157} 129

\bibitem{fhrouda02a} Hrouda F 2002 Low-field variation of magnetic susceptibility and its effect
on the anisotropy of magnetic susceptibility of rocks {\it Geophys. J. Int.} {\bf 150} 715

\end{thebibliography}
\end{document}